\documentclass[a4paper]{jpconf}
\usepackage{graphicx}
\newcommand{\ba}{\begin{eqnarray}}
\newcommand{\ea}{\end{eqnarray}}

\begin{document}

\title{Pentaquark states with hidden charm}

\author{Roelof Bijker}
\address{Instituto de Ciencias Nucleares, 
Universidad Nacional Aut\'onoma de M\'exico, 
A.P. 70-543, 04510 M\'exico, D.F., M\'exico}
\ead{bijker@nucleares.unam.mx}

\begin{abstract}
I develop an extension of the usual three-flavor quark model to four flavors 
($u$, $d$, $s$ and $c$), and discuss the classification of pentaquark 
states with hidden charm. This work is motivated by the recent observation of 
such states by the LHCb Collatoration at CERN. 
\end{abstract}

\section{Introduction}

The recent observation of candidates for pentaquark states by the LHCb Collaboration 
\cite{LHCb1,LHCb2,LHCb3} has generated an enormous renewed interest in these multiquark 
states with almost weekly publications in the arXive and the scientific literature. 
The first evidence was found in the $\Lambda_b^0 \rightarrow J/\psi p K^-$ decay \cite{LHCb1}. 
An analysis of the $J/\psi$ invariant mass spectrum showed the need to introduce two 
hidden charm pentaquark states, $P_c^+(4380)$ and $P_c^+(4450)$, with opposite parity. 
The evidence for the existence of these pentaquark states was confirmed later in a 
model-independent analysis \cite{LHCb2}, as well as in the 
$\Lambda_b^0 \rightarrow J/\psi p \pi^-$ decay \cite{LHCb3}. 
So far, the angular momentum and parity could not be determined uniquely. The most likely 
values are $J^P=3/2^-$ for $P_c^+(4380)$, and $5/2^+$ for $P_c^+(4450)$, 
although other combinations, like ($3/2^+$, $5/2^-$) or  
($5/2^+$, $3/2^-$), are not excluded. There are proposals to study the 
$P_c^+(4380)$ and $P_c^+(4450)$ states in photoproduction of $J/\psi$ on the proton 
to verify their existence and to determine their spin and parity \cite{photo1}-\cite{JLab}. 

Theoretically, several narrow $N^*$ and $\Lambda^*$ resonances with hidden charm were predicted 
above 4 GeV in a coupled-channel unitary approach \cite{Oset1,Oset2} several years before the LHCb 
data, as well as \cite{YangZC,Karliner}. 
After the discovery of the $P_c^+$ pentaquark states many different explanations have been 
explored, ranging from weakly bound molecular states of charmed baryons and mesons 
\cite{Karliner}-\cite{Yamaguchi}, and interpretations in terms of multiquark degrees of freedom 
\cite{Maiani}-\cite{Elena}, to kinematic effects \cite{nonresonant1,nonresonant2,nonresonant3}. 
More information on the experimental and theoretical aspects of pentaquark states,  
as well as a more complete list of references, can be found in the reviews \cite{Burns,Chen,PDG}. 

It is the aim of the present contribution to present a classification of all possible 
$uudc\bar{c}$ hidden-charm pentaquark states in an extension of the three- to the 
four-flavor $SU(4)$ quark model, and to calculate the corresponding magnetic moments 
of these configurations. This work is an extension of earlier studies of $uudd\bar{s}$ 
pentaquarks \cite{BGS1,BGS2}.

\section{Four-flavor SU(4) quark model} 

In this section, I review the four-flavor $SU(4)$ quark model which is based on the 
flavors up, down, strange and charm \cite{Bjorken}-\cite{DeRujula}. Even though 
the $SU(4)$ flavor symmetry is expected to be much more broken than the three-flavor 
$SU(3)$ flavor symmetry due to the large mass of the charm quark, it nevertheless 
provides a useful classification scheme for baryons and mesons.  

The spin-flavor states of the $udsc$ quark model can be decomposed into their flavor 
and spin parts as 
\ba
\left| \begin{array}{cccccc}
& SU_{\rm sf}(8) &\supset& SU_{\rm f}(4) &\otimes& SU_{\rm s}(2) \\
& [f] &,& [g] &,& S 
\end{array} \right> ~,
\ea
followed by a reduction to the three-flavor $uds$ states 
\ba
\left| \begin{array}{cccccc}
& SU_{\rm f}(4) &\supset& SU_{\rm f}(3) &\otimes& U_{\rm Z}(1) \\
& [g] &,& [h] &,& Z 
\end{array} \right> ~,
\ea
and finally to their isospin and hypercharge contents  
\ba
\left| \begin{array}{cccccc}
& SU_{\rm f}(3) &\supset& SU_{\rm I}(2) &\otimes& U_{\rm Y}(1) \\
& [h] &,& I &,& Y
\end{array} \right> ~.
\ea
The quarks transform as the fundamental representation $[1]$ under $SU(n)$, where $n=2$ 
for the spin, $n=3$ for the color, $n=4$ for the flavor, and $n=8$ for the combined spin-flavor 
degrees of freedom, whereas the antiquarks transform as the conjugate representation $[1^{n-1}]$ 
under $SU(n)$. The hypercharges $Y$ and $Z$ are related to the baryon number $B$, the strangeness 
${\cal S}$ and charm $C$ by \cite{PDG}
\ba
Y &=& B+{\cal S}-\frac{C}{3} ~,
\nonumber\\
Z &=& \frac{3}{4}B-C ~,
\ea
and the electric charge $Q$ is given by the generalized Gell-Mann-Nishijima relation 
\ba
Q &=& I_3+\frac{B+{\cal S}+C}{2} ~.
\ea

\section{$qqqq\bar{q}$ Pentaquark states}

The pentaquark wave functions depend both on the spatial degrees of 
freedom and the internal degrees of freedom of color, flavor and spin  
\ba
\psi = \psi^{\rm o} \phi^{\rm f} \chi^{\rm s} \psi^{\rm c} ~.
\ea
In order to classify the corresponding states, we shall make use as much as possible 
of symmetry principles. In the construction of the classification scheme we are guided 
by two conditions: (i) the pentaquark wave function should be antisymmetric under any 
permutation of the four quarks, and (ii) as all physical states, it should be a color singlet. 
In addition, we restrict ourselves to flavor multiplets containing $uudc\bar{c}$ 
configurations without orbital excitations and with spin and parity $J^P=S^P=3/2^-$. 

The permutation symmetry among the four quarks is characterized by the $S_4$ Young 
tableaux $[4]$, $[31]$, $[22]$, $[211]$ and $[1111]$ or, equivalently, by the irreducible 
representations of the tetrahedral group ${\cal T}_d$ (which is isomorphic 
to the permutation group ${\cal T}_d \sim S_4$) as $A_1$, $F_2$, $E$, $F_1$ and $A_2$, respectively.

\begin{table}
\centering
\caption[]{Spin-flavor decomposition of $q^4$ states. 
The subindices denote the dimension of the multiplet.}
\label{sfqqqq}
\begin{tabular}{ccccccc}
\noalign{\smallskip}
\hline
\noalign{\smallskip}
$SU_{\rm sf}(8)$ &$\supset$& $SU_{\rm f}(4)$ 
&$\otimes$& $SU_{\rm s}(2)$ & \\
\noalign{\smallskip}
$[f]$ &$\supset$& $[g]$ &$\otimes$& $[g']$ & $\psi^{\rm sf}_{F_2}$ \\
\noalign{\smallskip}
\noalign{\smallskip}
\hline
\noalign{\smallskip}
$[31]_{630}$ & & $[4]_{35}$   & $\otimes$ & $[31]_{3}$ & $\left[ \phi_{A_1} \times \chi_{F_2} \right]_{F_2}$ \\
             & & $[31]_{45}$  & $\otimes$ & $[4]_{5}$  & $\left[ \phi_{F_2} \times \chi_{A_1} \right]_{F_2}$ \\
             & &              &           & $[31]_{3}$ & $\left[ \phi_{F_2} \times \chi_{F_2} \right]_{F_2}$ \\ 
             & &              &           & $[22]_{1}$ & \\ 
             & & $[22]_{20}$  & $\otimes$ & $[31]_{3}$ & $\left[ \phi_{E}   \times \chi_{F_2} \right]_{F_2}$ \\
             & & $[211]_{15}$ & $\otimes$ & $[31]_{3}$ & $\left[ \phi_{F_1} \times \chi_{F_2} \right]_{F_2}$ \\ 
             & &              &           & $[22]_{1}$ & \\ 
\noalign{\smallskip}
\hline
\end{tabular}
\end{table}

The total pentaquark wave function has to be a $[222]$ color-singlet state. 
Since the color wave function of the antiquark is a $[11]$ anti-triplet, 
the color wave function of the four-quark configuration has to be a $[211]$ 
triplet with $F_1$ symmetry under ${\cal T}_d$. 
The total $q^4$ wave function is antisymmetric ($A_2$), hence the 
orbital-spin-flavor part is a $[31]$ state with $F_2$ symmetry 
\ba
\psi &=& \left[ \psi^{\rm c}_{F_1} \times \psi^{\rm osf}_{F_2} \right]_{A_2} ~, 
\ea
where the subindices refer to the symmetry properties of the four-quark system under permutation. 
Moreover, it is assumed that the orbital part of the pentaquark wave function 
corresponds to the ground state (without orbital excitations) with $A_1$ symmetry. 
As a consequence the spin-flavor part of the four-quark wave function has $F_2$ symmetry  
\ba
\psi^{\rm osf}_{F_2} &=& \left[ \psi^{\rm o}_{A_1} \times \psi^{\rm sf}_{F_2} \right]_{F_2}  ~. 
\ea
The flavor and spin content of the only allowed spin-flavor configuration, $[31]$ with 
$F_2$ symmetry, is given in Table~\ref{sfqqqq}. There are four different flavor multiplets 
$[g]=[4]$, $[31]$, $[22]$ and $[211]$. For pentaquark states with $J^P=3/2^-$ without orbital 
excitations the allowed values of the spin of the four-quark system are $S=(g'_1-g'_2)/2=1$ 
and $2$. In the last column, we show the allowed spin-flavor configurations, where $\phi$ and 
$\chi$ denote the flavor and spin wave functions, respectively. 

\begin{table}
\centering
\caption[]{$SU_{\rm f}(4) \supset SU_{\rm f}(3) \otimes U_{\rm Z}(1)$ flavor classification of 
four-quark states (here $q=u$, $d$, $s$).}
\label{fqqqq}
\begin{tabular}{ccccccccccc}
\noalign{\smallskip}
\hline
\noalign{\smallskip}
$SU_{\rm f}(4)$ &$\supset$& $SU_{\rm f}(3)$ && && && && \\
\noalign{\smallskip}
$[g]$ &$\supset$& $[h]$ && && && && \\
\noalign{\smallskip}
\hline
\noalign{\smallskip}
$[4]_{35}$   &$\supset$& $[4]_{15}$ &$\oplus$& $[3]_{10}$ &$\oplus$& $[2]_{6}$ &$\oplus$& $[1]_{3}$ &$\oplus$& $[0]_{1}$ \\
\noalign{\smallskip}
$[31]_{45}$  &$\supset$& $[31]_{15}$ &$\oplus$& $[3]_{10} \oplus [21]_{8}$ &$\oplus$& $[2]_{6} \oplus [11]_{3}$ &$\oplus$& $[1]_{3}$ && \\
\noalign{\smallskip}
$[22]_{20}$  &$\supset$& $[22]_{6}$ &$\oplus$& $[21]_{8}$ &$\oplus$& $[2]_{6}$ && && \\
\noalign{\smallskip}
$[211]_{15}$ &$\supset$& $[211]_{3}$ &$\oplus$& $[21]_{8} \oplus [111]_{1}$ &$\oplus$& $[11]_{3}$ && && \\
\noalign{\smallskip}
\hline
\noalign{\smallskip}
&& $qqqq$ && $qqqc$ && $qqcc$ && $qccc$ && $cccc$ \\
&& $Z= 1$ && $Z= 0$ && $Z=-1$ && $Z=-2$ && $Z=-3$ \\ 
\noalign{\smallskip}
\hline
\end{tabular}
\end{table}

Since the current interest is in four-quark states with one charm quark, it is convenient to 
make a decomposition of four into three flavors according to 
$SU_{\rm f}(4) \supset SU_{\rm f}(3) \otimes U_{\rm Z}(1)$. 
Table~\ref{fqqqq} shows that the states with one charm quark have $Z=0$ and belong to 
either an $SU(3)$ decuplet ($[h]=[3]$), octet ($[21]$) or singlet ($[111]$). Since the flavor singlet 
corresponds to a $udsc$ configuration, the only $SU(3)$ flavor multiplets that contain a $uudc$ state 
are the decuplet and the octet. 

\begin{figure}
\centering
\setlength{\unitlength}{0.6pt}
\begin{picture}(650,250)(0,0)
\thicklines
\put(230,215){$uudc\bar{c}$}
\put( 50,200) {\line(1,0){300}}
\put(100,150) {\line(1,0){200}}
\put(150,100) {\line(1,0){100}}
\put(200, 50) {\line(1,1){150}}
\put(150,100) {\line(1,1){100}}
\put(100,150) {\line(1,1){ 50}}
\put(200, 50) {\line(-1,1){150}}
\put(250,100) {\line(-1,1){100}}
\put(300,150) {\line(-1,1){ 50}}
\put(250,200){\circle*{10}}
\multiput( 50,200)(100,0){4}{\circle*{5}}
\multiput(100,150)(100,0){3}{\circle*{5}}
\multiput(150,100)(100,0){2}{\circle*{5}}
\put(200, 50){\circle*{5}}
\put(530,215){$uudc\bar{c}$}
\put(450,200) {\line(1,0){100}}
\put(400,150) {\line(1,0){200}}
\put(450,100) {\line(1,0){100}}
\put(400,150) {\line(1,1){ 50}}
\put(450,100) {\line(1,1){100}}
\put(550,100) {\line(1,1){ 50}}
\put(450,100) {\line(-1,1){ 50}}
\put(550,100) {\line(-1,1){100}}
\put(600,150) {\line(-1,1){ 50}}
\put(550,200){\circle*{10}}
\multiput(450,200)(100,0){2}{\circle*{5}}
\multiput(4000,150)(100,0){3}{\circle*{5}}
\multiput(450,100)(100,0){2}{\circle*{5}}
\put(500,150){\circle{10}}
\end{picture}
\caption{Pentaquark decuplet and octet}
\label{decuplet}
\end{figure}
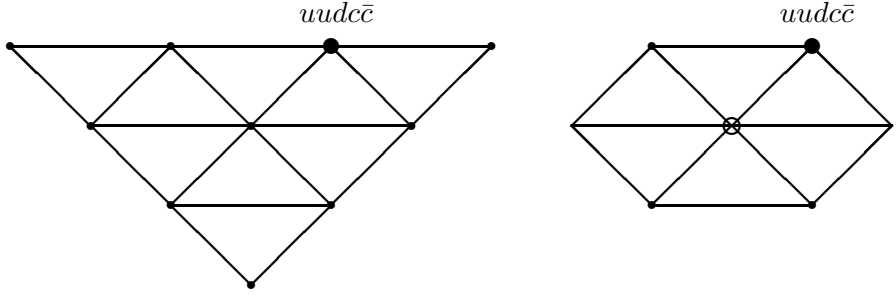

Finally, the $uudc\bar{c}$ pentaquark states belong to $SU(3)$ multiplets that are  
obtained by combining the four-quark $[3]$ decuplet and $[21]$ octet with the $[111]$ 
singlet for the antiquark with charm $\bar{c}$ 
\ba
\, [3] \otimes [111] &=& [411] ~,
\nonumber\\
\, [21] \otimes [111] &=& [321] ~. 
\ea
In summary, the $uudc\bar{c}$ pentaquark states belong to a $SU(3)$ flavor decuplet with 
$[h]=[411] \equiv [3]$ or an octet $[321] \equiv [21]$ (see Fig.~\ref{decuplet}). 
The corresponding flavor wave functions can be expressed in terms of the quantum numbers  
\ba
\left| \phi_{A_1} \right> &=& \left| [g],[h]_{qqqq},[h]_{\bar{q}};[h],I,I_3,Y,Z \right> ~,
\nonumber\\ 
&=& \left| [4],[3],[111];[411],\frac{3}{2},\frac{1}{2},1,\frac{3}{4} \right> ~,
\nonumber\\
\left| \phi_{F_2} \right> &=& \left| [31],[3],[111];[411],\frac{3}{2},\frac{1}{2},1,\frac{3}{4} \right> ~, 
\ea
for the decuplet, and  
\ba
\left| \phi_{F_2} \right> &=& \left| [31],[21],[111];[321],\frac{1}{2},\frac{1}{2},1,\frac{3}{4} \right> ~,
\nonumber\\
\left| \phi_{E}   \right> &=& \left| [22],[21],[111];[321],\frac{1}{2},\frac{1}{2},1,\frac{3}{4} \right> ~,
\nonumber\\
\left| \phi_{F_1} \right> &=& \left| [211],[21],[111];[321],\frac{1}{2},\frac{1}{2},1,\frac{3}{4} \right> ~,
\ea
for the octet. 

\section{Magnetic moments}

In this section, we study the magnetic moments of the $uudc\bar{c}$ pentaquark states 
with spin and parity $J^P=3/2^-$ discussed in the previous section. The magnetic moment 
is a crucial ingredient in calculation of the photoproduction cross sections 
of pentaquarks. In Ref.~\cite{Wang}, the magnetic moments were calculated for a molecular 
model, a diquark-triquark model and a diquark-diquark-antiquark model of pentaquark 
configurations. To the best of our knowledge, the present calculation is the first for a  
constituent quark model of pentaquarks. 

The magnetic moment of a multiquark system is given by the sum of the magnetic moments of its 
constituent parts \cite{BGS2}
\ba
\vec{\mu} = \vec{\mu}_{\rm spin} + \vec{\mu}_{\rm orb} = \sum_i \mu_i(2\vec{s}_i + \vec{l}_i) ~,
\ea
where $\mu_i$ denotes the magnetic moment of the $i-$th quark or antiquark.

Since at present, we do not consider radial or orbital excitations, the magnetic moment only 
depends on the spin part. The results for the different $uudc\bar{c}$ configurations are given 
in Table~\ref{magmom}. The numerical values are obtained by using
\ba
\mu_u &=& +1.852 \;\mu_N ~,
\nonumber\\
\mu_d &=& -0.972 \;\mu_N ~,
\nonumber\\
\mu_s &=& -0.613 \;\mu_N ~,
\nonumber\\
\mu_c &=& +0.404 \;\mu_N ~.
\ea
The magnetic moments of the up, down and strange quarks were fitted to the magnetic moments of 
the proton, neutron and $\Lambda$ hyperon. The value of the magnetic moment of the charm quark 
is consistent with a constituent mass of 1.550 GeV. 

\begin{table}
\centering
\caption{Magnetic moments of $uudc\bar{c}$ pentaquark states in $\mu_N$.}
\label{magmom}
\begin{tabular}{lclr}
\noalign{\smallskip}
\hline
\noalign{\smallskip}
& State & \multicolumn{2}{c}{Magnetic moment} \\
\noalign{\smallskip}
\hline
\noalign{\smallskip}
Decuplet & $\left[ \phi_{A_1} \times \chi_{F_2} \right]_{F_2}$ & $\frac{1}{2}(2\mu_u+\mu_d+\mu_c)+\mu_{\bar{c}}$ & 1.164 \\
\noalign{\smallskip}
& $\left[ \phi_{F_2} \times \chi_{A_1} \right]_{F_2}$ & $\frac{9}{10}(2\mu_u+\mu_d+\mu_c)-\frac{3}{5}\mu_{\bar{c}}$ & 3.065 \\
\noalign{\smallskip}
& $\left[ \phi_{F_2} \times \chi_{F_2} \right]_{F_2}$ & $\frac{2}{3}(2\mu_u+\mu_d)+\mu_{\bar{c}}$ & 1.417 \\
\noalign{\smallskip}
\mbox{Octet} & $\left[ \phi_{F_2} \times \chi_{A_1} \right]_{F_2}$ & $\frac{9}{10}(2\mu_u+\mu_d+\mu_c)-\frac{3}{5}\mu_{\bar{c}}$ & 3.065 \\
\noalign{\smallskip}
& $\left[ \phi_{F_2} \times \chi_{F_2} \right]_{F_2}$ & $\frac{1}{12}(14\mu_u+\mu_d+9\mu_c)+\mu_{\bar{c}}$ & 1.979 \\
\noalign{\smallskip}
& $\left[ \phi_{E} \times \chi_{F_2} \right]_{F_2}$ & $\frac{1}{2}(2\mu_u+\mu_d+\mu_c)+\mu_{\bar{c}}$ & 1.164 \\
\noalign{\smallskip}
& $\left[ \phi_{F_1} \times \chi_{F_2} \right]_{F_2}$ & $\frac{1}{4}(6\mu_u+\mu_d+\mu_c)+\mu_{\bar{c}}$ & 2.232 \\
\noalign{\smallskip}
\hline
\end{tabular}
\end{table}

\section{Summary and conclusions}

In conclusion, in this contribution we discussed the classification of $uudc\bar{c}$ pentaquark states 
with angular momentum and parity $J^P=3/2^-$. At present we did not include orbital excitations. The coupling 
between the four-quark states $uudc$ and the antiquark $\bar{c}$ was carried out at the level of the $SU(3)$ 
flavor group which results in a pure $uudc\bar{c}$ state. Couplings at the level of the $SU(8)$ spin-flavor 
group or the $SU(4)$ flavor group will lead to admixtures with $uudu\bar{u}$, $uudd\bar{d}$ and $uuds\bar{s}$ 
configurations \cite{Bijker}. 

Since for future studies of pentaquark states in photoproduction reactions it is important to have 
information on the electromagnetic couplings, we presented a first step by calculating the magnetic moments 
of all allowed $uudc\bar{c}$ states. The numerical values are positive and in the range from $\sim 1-3$ $\mu_N$. 

Finally, if indeed the pentaquark exists, there should be an entire multiplet of pentaquark states. The study 
of the structure of these multiplets and the mass spectrum is the subject of work in progress \cite{Bijker}.

\ack

This work was supported in part by grant IN109017 from PAPIIT-DGAPA, UNAM 
and grant 251817 from CONACyT, Mexico.


\section*{References}

\begin{thebibliography}{99} 

\bibitem{LHCb1}
Aaij R {\it et al.} (LHCb Collaboration) 2015 {\it Phys. Rev. Lett.} {\bf 115} 072001

\bibitem{LHCb2}
Aaij R {\it et al.} (LHCb Collaboration) 2016 {\it Phys. Rev. Lett.} {\bf 117} 082002

\bibitem{LHCb3}
Aaij R {\it et al.} (LHCb Collaboration) 2016 {\it Phys. Rev. Lett.} {\bf 117} 082003

\bibitem{photo1}
Kubarovsky V and Voloshin M B 2015 {\it Phys. Rev. D} {\bf 92} 031502(R) 

\bibitem{photo2}
Wang Q, Liu X H and Zhao Q 2015 {\it Phys. Rev. D} {\bf 92} 034022 

\bibitem{photo3}
Karliner M and Rosner J L 2016 {\it Phys. Lett. B} {\bf 752} 329 

\bibitem{photo4}
Hiller Blin A N, Fern\'andez-Ram{\'{\i}}rez C, Jackura A, Mathieu V, Mokeev V I, 
Pilloni A and Szczepaniak A P 2016 {\it Phys. Rev. D} {\bf 94} 034002 

\bibitem{photo5}
Fern\'andez-Ram{\'{\i}}rez C, Hiller Blin A N and Pilloni A 2017 {\it arXiv:1703.06928}

\bibitem{JLab}
Meziani Z E {\it et al.} 2016 {\it arXiv:1609.00676}

\bibitem{Oset1}
Wu J J, Molina R, Oset E and Zou B S 2010 {\it Phys. Rev. Lett.} {\bf 105} 232001

\bibitem{Oset2}
Wu J J, Molina R, Oset E and Zou B S 2011 {\it Phys. Rev. C} {\bf 84} 015202 

\bibitem{YangZC}
Yang Z C, Sun Z F, He J, Liu X and Zhu S L 2012 {\it Chin. Phys. C} {\bf 36} 6

\bibitem{Karliner}
Karliner M and Rosner J L 2015 {\it Phys. Rev. Lett.} {\bf 115} 122001

\bibitem{RuiChen}
Chen R, Liu X, Li X Q and Zhu S L 2015 {\it Phys. Rev. Lett.} {\bf 115} 132002

\bibitem{QCDSR}
Chen H X, Chen W, Liu X, Steele T G and Zhu S L 2015 {\it Phys. Rev. Lett.} {\bf 115} 172001

\bibitem{Roca}
Roca L, Nieves J and Oset E 2015 {\it Phys. Rev. D} {\bf 92} 094003

\bibitem{He}
He J 2016 {\it Phys. Lett. B} {\bf 753} 547

\bibitem{Eides}
Eides M I, Petrov V Yu, Polyakov M V 2016 {\it Phys. Rev. D} {\bf 93} 054039

\bibitem{Yamaguchi}
Yamaguchi Y and Santopinto E 2016 {\it arXiv:1606.08330}

\bibitem{Maiani}
Maiani L, Polosa A D and Riquer V 2015 {\it Phys. Lett. B} {\bf 749} 289

\bibitem{Lebed}
Lebed R F 2015 {\it Phys. Lett. B} {\bf 749} 454

\bibitem{Wang}
Wang G J, Chen R, Ma L, Liu X and Zhu S L 2016 {\it Phys. Rev. D} {\bf 94} 094018

\bibitem{Yang}
Yang G, Ping J and Wang F 2017 {\bf Phys. Rev. D} {\bf 95} 014010

\bibitem{Deng}
Deng C, Ping J, Huang H and Wang F 2017 {\it Phys. Rev. D} {\bf 95} 014031	

\bibitem{Takeuchi}
Takeuchi S and Takizawa M 2017 {\it Phys. Lett. B} {\bf 764} 254

\bibitem{Elena}
Santopinto E and Giachino A 2017 {\it arXiv:1604.03769v2}

\bibitem{nonresonant1}
Guo F K, Meissner U G, Wang W and Yang Z 2015 {\it Phys. Rev. D} {\bf 92} 071502

\bibitem{nonresonant2}
Liu X H, Wang Q and Zhao Q 2016 {\it Phys. Lett. B} {\bf 757} 231

\bibitem{nonresonant3}
Mikhasenko M 2015 arXiv:1507.06552 

\bibitem{Burns}
Burns T J 2015 {\it Eur. Phys. J. A} {\bf 51} 152

\bibitem{Chen}
Chen H X, Chen W, Liu X and Zhu S L 2016 {\it Phys. Rep.} {\bf 639} 1

\bibitem{PDG}
Patrignani C {\it et al.} (Particle Data Group) 2016 {\it Chin. Phys. C} {\bf 40} 100001

\bibitem{BGS1}
Bijker R, Giannini M M and Santopinto E 2004 {\it Eur. Phys. J. A} {\bf 22} 319 

\bibitem{BGS2}
Bijker R, Giannini M M and Santopinto E 2004 {\it Phys. Lett. B} {\bf 595} 260 

\bibitem{Bjorken}
Bj\o{}rken B J and Glashow S L 1964 {\it Phys. Lett.} {\bf 11} 255

\bibitem{GIM}
Glashow S L, Iliopoulos J and Maiani L 1970 {\it Phys. Rev. D} {\bf 2} 1285

\bibitem{charm}
Gaillard M K, Lee B W and Rosner J L 1975 {\it Rev. Mod. Phys.} {\bf 47} 277 

\bibitem{DeRujula}
De R\'ujula A, Georgi H and Glashow S L 1975 {\it Phys. Rev. D} {\bf 12} 147

\bibitem{Bijker}
Bijker R and Fern\'andez-Ram{\'{\i}}rez C 2017, work in progress

\end{thebibliography}
\end{document}